\documentclass[tradiabstract]{aa} % for the abstract without structuration 

\usepackage{graphicx}
\usepackage{txfonts}

\usepackage{epsf}
\usepackage{rotating}
\usepackage{graphics}

\def \MJ{M$_{\mathrm{Jup}}$}

\def \kms{km\,s$^{-1}$}
\def \ms{m\,s$^{-1}$}
\def \1s{$1\,\sigma$}
\def \kid{$\chi^2$}
\def \t0{T$_0$}

\def \cible{BD\,$-08^{\circ}2823$}
\def \cibleb{{\cible}b}
\def \ciblec{{\cible}c}
\def \cibled{{\cible}d}
\def \hip{{\it Hipparcos}}

\begin{document}

   \title{The HARPS search for southern extra-solar planets\thanks{
Based on observations made with HARPS spectrograph on the 3.6-m 
ESO telescope at La Silla Observatory, Chile, under the programs ID 
072.C-0488, 074.C-0364 and 078.C-0044.}}
            
   \subtitle{XX. Planets around the active star \cible}

   \author{G.~H\'ebrard\inst{1}, 
   	  S.~Udry\inst{2},
	  G.~Lo~Curto\inst{3}, 
	  N.~Robichon\inst{4},
	  D.~Naef\inst{2, 5},
	  D.~Ehrenreich\inst{1,6},
	  W.~Benz\inst{7},
	  F.~Bouchy\inst{1,8},
	  A.~Lecavelier des Etangs\inst{1}, 
	  C.~Lovis\inst{2},
	  M.~Mayor\inst{2},
	  C.~Moutou\inst{9},
	  F.~Pepe\inst{2},
	  D.~Queloz\inst{2},
	  N.~C.~Santos\inst{10},
           D.~S\'egransan\inst{3}
}

   \institute{
%1
Institut d'Astrophysique de Paris, UMR7095 CNRS, Universit\'e Pierre \& Marie Curie, 
98bis boulevard Arago, 75014 Paris, France 
\and
%2
Observatoire de Gen\`eve, Universit\'e de Gen\`eve, 51 Chemin des Maillettes, 1290 Sauverny, Switzerland
\and
%3
ESO, Karl-Schwarzschild-Strasse 2, D-85748 Garching bei M\"unchen, Germany 
\and
%4
Observatoire de Paris, GEPI, 5 Place Jules Janssen, F-92195 Meudon, France
\and
%5
ESO, Alonso de Cordoba 3107, Vitacura Casilla 19001, Santiago, Chile 
\and
%6
Laboratoire d'Astrophysique de Grenoble,   %Observatoire de Grenoble, 
CNRS (UMR 5571), Universit\'e J. Fourier, BP53, 38041 Grenoble cedex 9, France
\and
%7
Physikalisches Institut Universit\"at Bern, Sidlerstrasse 5, 3012 Bern, Switzerland 
\and
%8
Observatoire de Haute-Provence, CNRS/OAMP, 04870 Saint-Michel-l'Observatoire, France
\and
%9
Laboratoire d'Astrophysique de Marseille, Universit\'e de Provence, CNRS (UMR 6110), 
BP 8, 13376 Marseille Cedex 12, France
\and
Centro de Astrof{\'\i}sica, Universidade do Porto, Rua das Estrelas, 4150-762 Porto, Portugal
}

   \date{Received TBC; accepted TBC}
      
\abstract{We report the detection of a planetary system around \cible, that includes at least one 
Uranus-mass planet and one Saturn-mass planet. This discovery serendipitously originates from 
a search for planetary transits in the \hip\ photometry database. This program preferentially selected 
active stars and did not allow the detection of new transiting planets. 
It allowed however the identification of the K3V star \cible\ as a target 
harboring a multiplanet system, that we secured and 
characterized thanks to an intensive monitoring with the HARPS spectrograph at the 3.6-m 
ESO telescope in La Silla. The stellar activity level of \cible\ complicates the analysis but does 
not prohibit the detection of two planets around this star. 
\cibleb\ has a minimum mass of $14.4\pm2.1$~M$_{\oplus}$ and an orbital period of 
5.60~days, whereas \ciblec\ has a minimum mass of $0.33\pm0.03$~\MJ\ and an orbital 
period of 237.6~days. 
This new system strengthens the fact that low-mass planets are preferentially found in
multiplanetary systems, but not around high-metallicity stars as this is the case for 
massive planets. It also supports the belief  
that active stars should not be neglected in exoplanet searches, even when searching 
for~low-mass~planets.
}


\authorrunning{H\'ebrard et al.}
\titlerunning{Planets around the active star \cible}

\maketitle

%________________________________________________________________

\section{Introduction}
\label{sect_intro}

The HARPS spectrograph (Mayor et al.~\cite{mayor03})
is operating since 2003 at the 3.6-m ESO telescope in La Silla, Chile.
This is a fiber-fed, environmentally stabilized high-resolution echelle spectrograph 
dedicated to high-precision radial velocity measurements. Thanks to the 
exoplanetology programs that are conducted with it, it allowed numerous 
extra-solar planets studies and discoveries, in the ranges of 
Jupiter-mass planets (e.g. Pepe et al.~\cite{pepe04}, Moutou et al.~\cite{moutou09a}), 
low-mass planets (e.g. Lovis et al.~\cite{lovis06}, Mayor et al.~\cite{mayor09}), 
planets around early- (e.g. Desort et al.~\cite{desort08}, Lagrange et al.~\cite{lagrange09})
or late-type stars (e.g. Bonfils et al.~\cite{bonfils07}, Forveille et al.~\cite{forveille09}),
or transiting planets (e.g. Bouchy et al..~\cite{bouchy08}, Queloz et al.~\cite{queloz09}).
The essential quality of HARPS (the High Accuracy Radial 
velocity Planet Searcher) is its high stability, which results in a sub-\ms\ 
accuracy in the radial velocity measurements, on time-scales of several years. 
This allows the detection of planets in 
the Neptune or Super-Earth mass ranges, on progressively increasing orbital 
periods as the time-span of the monitoring is growing up.
This instrument hitherto~plays a major role in the improvement of the 
knowledges of~exoplanets.

We report here the detection of a new multiplanetary system. It serendipitously 
originates from the radial velocity follow-up accompanying a search 
for planetary transits in the \hip\ epoch photometry annex
(Perryman et al.~\cite{perryman97}). Despite its significant 
stellar activity, the star \cible\ was identified during this follow-up as a possible 
host of extra-solar planets. An intensive monitoring of its radial velocities was 
subsequently carried out with HARPS, that led to the detection of a planetary 
system around this star, including at least two planets.
We briefly present the search for planetary transits in the \hip\ photometry in 
\S~\ref{sect_hipparcos}. The HARPS observations of \cible\ are presented 
in \S~\ref{sect_stellar_properties} together with the stellar properties of this 
target. The planetary system is characterized in~\S~\ref{sect_planetary_system}, 
then the results are discussed~in~\S~\ref{sect_conclusion}.

\section{Search for transiting planets in the \hip\ database}
\label{sect_hipparcos}

The goal of our initial search in the \hip\ database
was to try to find new transiting planets, especially around bright stars.
Transiting planets could allow numerous important studies to be performed. 
These studies include
planetary radii, masses and densities measurements, accurate determination 
of the inclinations and the eccentricities of the orbits, possible detections of transit 
timing variations due to additional companions, measurements of the spin-orbit 
(mis)alignment angles, detections of the emission or absorption planetary 
spectra, or even satellites or rings detections. The famous planets transiting
in front of the bright stars HD\,209458 and HD\,189733 are those that allow 
the most accurate measurements and the deepest investigations.

The \hip\ epoch photometry annex contains between $\sim40$ and
$\sim300$ photometric measurements performed during the 1226-day duration 
of the mission for each of the 118\,204 stars of the catalog, in the
magnitude limit $V\le10$. 
With about 0.1\% chance that a given star harbors a transiting extra-solar 
planet, the \hip\ all-sky survey must contain photometric measurements for 
tens of transiting hot Jupiters. This belief is reinforced by the two cases of 
transiting planets {\it a posteriori} re-discovered in \hip\ data: 
HD\,209458b (Robichon \& Arenou~\cite{robichon00}, Castellano et 
al.~\cite{castellano00}, S\"oderhjelm~\cite{soderhjelm99}) and HD\,189733b 
(Bouchy et al.~\cite{bouchy05}, H\'ebrard \& Lecavelier des 
Etangs~\cite{hebrard06}). Thanks to the long time baseline, these data also 
allow the orbital period to be measured with an accuracy of the order 
of the second.

Up to now, there are the only two transiting planets that have been found 
in the \hip\ database. Indeed, with photometric variations of about 1\% or
less, such transits are difficult to identify in the \hip\ data, that
present a mean photometric accuracy of the same order of magnitude.
The poor time coverage, by comparison with dedicated surveys as 
SuperWASP, XO, CoRoT or Kepler, is another difficulty.
Jenkins et al.~(\cite{jenkins02}) concluded that because of its poor photometric
quality, the \hip\ catalog does not represent a likely place to detect
planets in the absence of other informations. It might however provide
planetary transit candidates for follow-up observations.
Laughlin~(\cite{laughlin00}) has searched in the \hip\ epoch photometry for
transiting planets within 206 metal-rich stars. None have been
confirmed~thereafter.

We present in Appendix~\ref{appendice_1} the systematic search we 
managed in the \hip\ epoch photometry annex for periods compatible 
with planetary transits. We constructed a ranked list of 
candidates for follow-up radial velocity measurements, based 
on the depth and significance of the planetary transits that could 
be detected in the \hip\ photometry. The simulations we performed 
indicated that our detection rate for transiting planets is low, 
of the order of 2~\%; so the chances to find new transiting planets 
in the \hip\ data are small (see Appendix~\ref{appendice_1}).
However, the particularly high interest for the potential discovery of 
a planet transiting a bright star pushed us to perform a radial velocity 
follow-up of our candidates.

We observed 194 of these selected, ranked targets with HARPS, 
in order to search for radial velocity variations in agreement with 
the transiting candidates found in the \hip\ photometry. The HARPS setup, 
the spectra extraction and the radial velocity measurements were identical 
to those described below in \S~\ref{sect_obs_BD08}. Most of these 
observations were performed in December~2004 as part of the 
program 074.C-0364. 
Radial velocity variations larger than 20~\ms\ were measured for 37 stars,
i.e. 19\% of our observed sample. Those variations are mainly caused by stellar 
activity. Indeed, most of the targets with large radial velocity variations show 
emissions in the \ion{Ca}{ii} H \& K  lines (3934.8\,\AA\ and 3969.6\,\AA). 
The level of these emissions is quantified with the activity S-index (Mount Wilson System), 
which is converted to the  $\log{R'_\mathrm{HK}}$ index (Santos et al.~\cite{santos00}). 
The majority of the variable targets have $\log{R'_\mathrm{HK}}$ indexes larger 
than $-4.7$, indicating prominent chromospheric activity; this can explain such level of radial 
velocity variations (Santos et al.~\cite{santos00}). 

Some targets first seemed to exhibit periodic radial velocity variations, but subsequent
monitoring with HARPS showed that these variations were not persistent with time, in 
agreement with transient activity processes on the stellar surface due to activity (flares, 
spots, plages...) modulated by the stellar rotation.  In addition, in most of the cases, 
analysis of the line profiles using bisectors of the cross-correlation functions shown 
variations in the shape of the lines with time. This 
indicates that the observed radial velocity variations are not 
due to Doppler shifts of the lines, but rather to deformations of the shape of the 
spectral lines.  Some cases revealed a clear anti-correlation between radial velocity and 
line-bisector orientation, which could be understood as  the signature of cool spots 
on the stellar photospheres (see, e.g., Queloz et al.~\cite{queloz01}, Melo et 
al.~\cite{melo07}, Desort et al.~\cite{desort07}, Boisse et al.~\cite{boisse09}).

Thus, our procedure seems to preferentially select active stars. This should be 
contrasted with transiting candidates obtained from the photometric surveys 
dedicated to transiting planet searches, which are mainly eclipsing binaries or 
transiting planets (see, e.g., Pont et al.~\cite{pont05}).
None of the targets that we observed through this program 
have shown radial velocity variations in agreement with a reflex motion
due to a hot Jupiter. Instead, most of the radial velocity variations seem to be
caused by stellar activity.

However, these observations allowed the serendipitously discovery of a 
new planetary system, without transit detection in the \hip\ data. This system 
orbits \cible, which is an active star, but among those with the lowest 
$\log{R'_\mathrm{HK}}$ indexes within our monitored targets. This activity 
level complicates planet detection but does not prohibit it in this case. 
We concentrate below on this target and the detection of its harbored 
planetary system.

\section{Observations and properties of \cible}
\label{sect_stellar_properties}

\subsection{HARPS observations of \cible}
\label{sect_obs_BD08}

We observed \cible\ using the HARPS  spectrograph 
at the 3.6-m ESO telescope in La Silla.
The bandpass of the spectra ranges from 3800~\AA\ to 6900~\AA, and the 
resolution power is $R=115\,000$, with a fiber diameter of 1\,arcsec. 
The spectra were extracted from the detector images with the HARPS pipeline, 
that includes localization of the orders on the 2D-images, optimal order extraction, 
cosmic-ray rejection, wavelength calibration and corrections of flat-field. 
The pipeline then performs a cross-correlation of the extracted spectra 
with a numerical mask (K5-type in this case), and finally measures the radial 
velocities from Gaussian fits of the cross-correlation functions (CCFs), 
following the method described by Baranne et al.~(\cite{baranne96}) 
and Pepe et al.~(\cite{pepe02}).
The full dataset we use for \cible\ includes 83 spectra. 
All the exposures (but the first one) were obtained without simultaneous thorium calibration.
The exposure times for \cible\ range between 4 and 15~minutes, allowing a 
signal-to-noise ratio per pixel between 30 and 80 to be reached around 550~nm.
The total exposure time is about 11~hours.

\begin{figure}[h] 
\begin{center}
\includegraphics[scale=0.44]{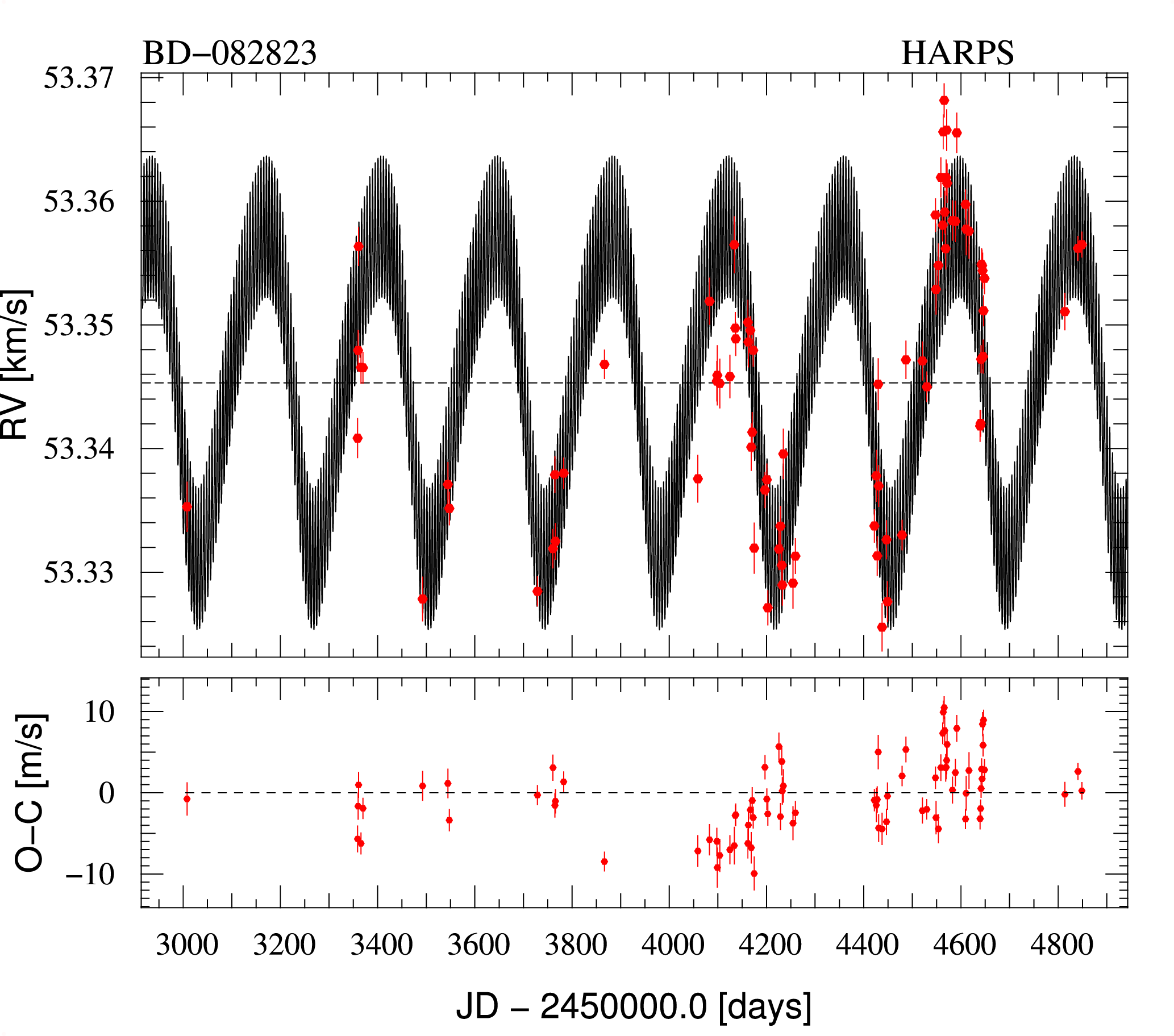}
\caption{\textit{Top:} Radial velocity HARPS measurements of \cible\ 
as a function of time (see \S~\ref{sect_stellar_properties}), 
and Keplerian fit with two planets (see \S~\ref{sect_planetary_system}). 
The orbital parameters corresponding to this 
fit are reported in Table~\ref{table_parameters}. 
\textit{Bottom:} Residuals of the fit with 1-$\sigma$~error bars.}
\label{fig_omc}
\end{center}
\end{figure}

The HARPS radial velocities of \cible\ as a function of time are plotted in the upper panel 
of  Fig.~\ref{fig_omc}. They span 1841~days, corresponding to 5.0~years.
A few measurements were obtained in late-2004 as part of the program 074.C-0364 
(see \S~\ref{sect_hipparcos}), then the target was regularly monitored. 
A few extra measurements were acquired in late-2006/early-2007 as part of the 
program 078.C-0044, but most of the data were secured as part of the 
\textit{Guaranteed Time Observations} (GTO) survey 
program 072.C-0488 (Mayor et al.~\cite{mayor03}). This is the case in particular 
for the intensive 
series in 2007-2008 (BDJ$-2\,400\,000$ between 54\,150 and 54\,650).
There was indeed an agreement to follow, as part of the GTO time, a few 
promising targets identified from the \hip\ original~program.

The derived radial velocities are reported in Table~\ref{table_rv}. The accuracies 
are between 1.0 and 2.4~\ms, typically around 1.6~\ms. This includes photon noise
but not  the jitter due to stellar oscillations or activity (\S~\ref{stellar_prop}). 
Wavelength calibration and spectrograph drift uncertainties 
are negligible with respect to the photon noise.
Four exposures with radial velocity uncertainties larger than 2.4~\ms\ were not 
included in the final dataset of 83~spectra.

The radial velocities show a significant 11-\ms~dispersion (43~\ms\ peak-to-peak), 
well over the expected accuracy (Fig.~\ref{fig_omc}, upper panel). The CCF from which 
those radial velocities were measured show parameters that also significantly vary 
with time. Their full widths at half maximum fluctuates between $6.22 \pm 0.01$ and 
$6.33\pm0.01$~\kms, and their contrasts from $46.80\pm0.05$ and $47.50\pm0.05$~\%\   
of the continuum. The bisector span of the CCF also show a dispersion, at the level 
of 7~\ms\ (see~Fig.~\ref{fig_bis}, upper panel). 
Thus, Doppler shifts of the spectral lines of \cible\ are not necessarily the only 
explanation for the observed radial velocity variations; changes in the shape of the lines, 
as those due to stellar activity, are likely to be the cause of at least a part of those~variations.

\begin{table}[h]
  \centering 
  \caption{HARPS measurements of \cible\ (full table available electronically)$^\dagger$.}
\label{table_rv}
\begin{tabular}{cccccc}
\hline
\hline
BJD & RV & $\sigma$(RV) & FWHM & Bis. & $\log{R'_\mathrm{HK}}$ \\
-2\,400\,000 & km\,s$^{-1}$  & km\,s$^{-1}$ & km\,s$^{-1}$ & km\,s$^{-1}$ &   \\
\hline
53007.83343   &   53.3353   &   0.0020   &  6.281 & 0.0222  & -4.702    \\   
53358.83576   &   53.3408   &   0.0016   &  6.248 & 0.0183  & -4.731    \\   
53359.80161   &   53.3479   &   0.0016   &  6.255 & 0.0154  & -4.725    \\   
53360.81961   &   53.3563   &   0.0015   &  6.252 & 0.0226  & -4.742    \\   
\ldots & \ldots & \ldots & \ldots & \ldots & \ldots \\
\ldots & \ldots & \ldots & \ldots & \ldots & \ldots \\
54648.45269   &   53.3538   &   0.0012   &  6.322 & 0.0326  & -4.709    \\   
54813.83365   &   53.3511   &   0.0015   &  6.291 & 0.0273  & -4.755    \\   
54840.78987   &   53.3562   &   0.0010   &  6.300 & 0.0294  & -4.732    \\   
54848.79967   &   53.3565   &   0.0010   &  6.283 & 0.0257  & -4.751    \\   
\hline
\multicolumn{6}{l}{$\dagger$: the six columns are respectively the Barycentric Julian Day of the} \\
\multicolumn{6}{l}{observations, the radial velocity and its uncertainty, the full width} \\
\multicolumn{6}{l}{at half maximum of the CCF, the span of its bisector, and the  \ion{Ca}{ii}} \\
\multicolumn{6}{l}{activity index.}
\end{tabular}
\end{table}

\subsection{Stellar properties and activity}
\label{stellar_prop}

Table~\ref{table_stellar_parameters} summarizes the stellar parameters of \cible\ 
(HIP\,49067, SAO\,137286). According to the SIMBAD database, this is 
K3V star of magnitude $V=9.86$. Its \hip\ parallax ($\pi= 23.76\pm1.61$~mas) 
implies a distance of $42.2 \pm 2.9 $~pc. The \hip\ color is $B-V=1.071 \pm 0.010$.
From spectral analysis of the HARPS data using the method presented in Santos et 
al.~(\cite{santos04a}), we derive the temperature $T_{\rm eff} = 4746 \pm 63 $~K, the 
gravity $\log g = 4.13 \pm 0.26 $, the metallicity 
${\rm [Fe/H]} = -0.07 \pm 0.03 $, and an uncertain age of $4.5 \pm 4.0$~Gyr.
The stellar mass we obtain is $M_* = 0.74 \,\rm{M}_{\odot}$, 
with a formal error bar of  $\pm 0.2 \,\rm{M}_{\odot}$. 
Following Fernandes \&\ Santos~(\cite{fernandes04}), 
we rather adopt a conservative $\pm10$~\%\ uncertainty, 
corresponding to $\pm0.07 \,\rm{M}_{\odot}$.
We derive a projected rotational velocity 
$v\sin i_\star = 1.4$~\kms\  from the parameters of the CCF using 
a calibration similar to that presented by Santos et al.~(\cite{santos02}). 
Acoustic oscillations are not averaged out in the data, as there are on 
timescales similar or shorter than the exposure times we used. However, 
for a K-type star, oscillation amplitudes are expected to be negligible by 
comparison with stellar activity effects on the measured radial~velocities.

\begin{table}[h]
  \centering 
\caption{Stellar parameters for \cible.}
  \label{table_stellar_parameters}
\begin{tabular}{lc}
\hline
\hline
Parameters  & Values \\ 
\hline
$m_v$                		&	$9.86$ 			\\ 
Spectral~type        		&	K3V				\\ 
$B-V$          			&	$1.071 \pm 0.010$ 	\\ 
Parallax [mas]			&	$23.76\pm 1.61 $	\\ 
Distance [pc]     		&	$42.2 \pm 2.9$ 		\\ 
$v\sin i_\star $ [\kms]		&	$1.4$			\\ 
$P_\mathrm{rot}$ [days]		&	$26.6 \pm 1.5$		\\ 
$\log{R'_\mathrm{HK}}$	&	$-4.8 \pm 0.1$		\\ 
${\rm [Fe/H]}$ 			&	$-0.07 \pm 0.03$	\\ 
$T_{\rm eff}$ [K]		&	$4746 \pm 63 $		\\ 
$\log g$ [cgs] 			&	$4.13 \pm 0.26$	\\ 
Mass~$[\rm{M}_{\odot}]$		&	$0.74 \pm 0.07$	\\ 
Age [Gyr]				&      $4.5 \pm 4.0$	\\ 
\hline
\end{tabular}
\end{table}

\begin{figure}[h] 
\begin{center}
\includegraphics[scale=0.64]{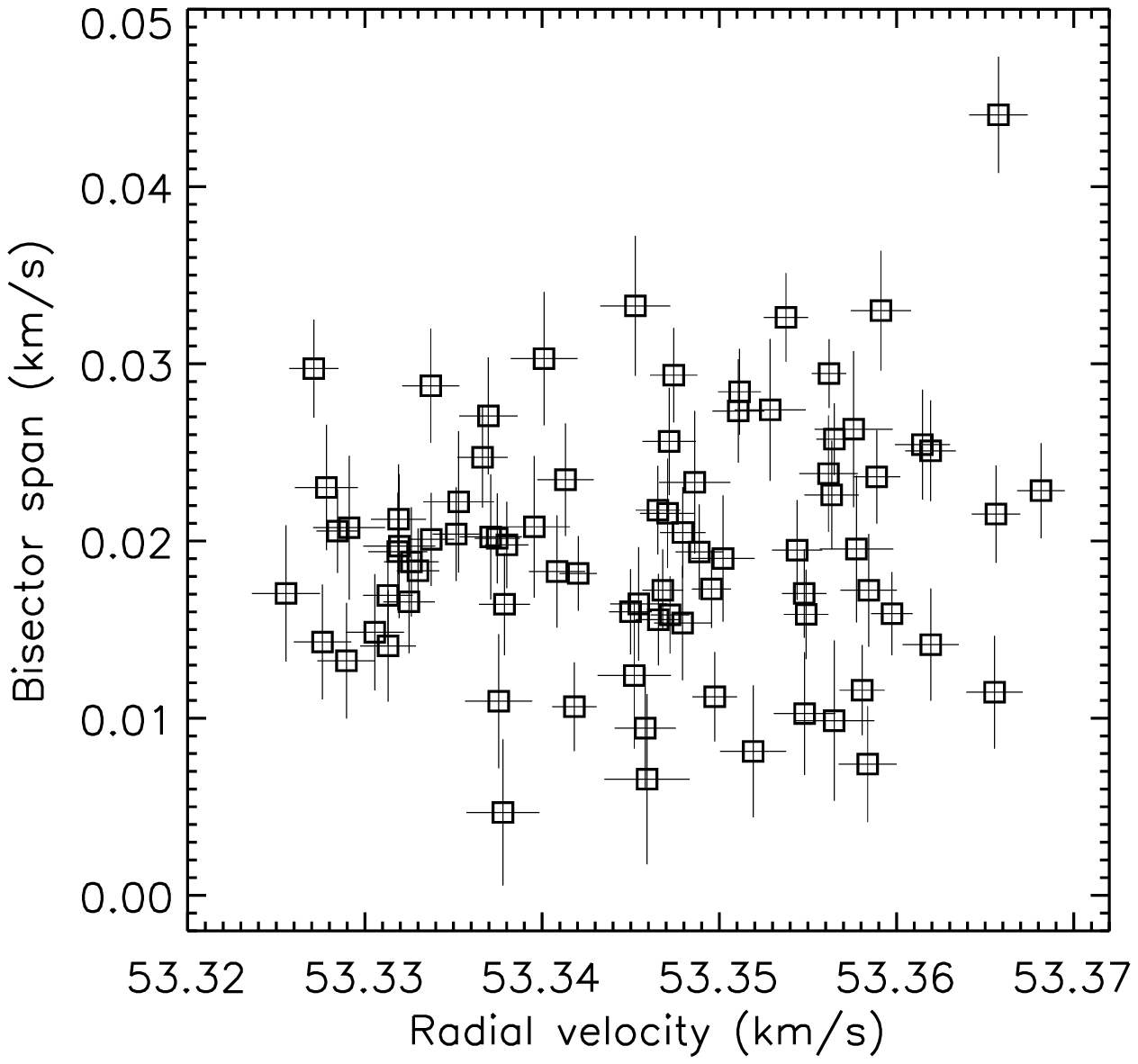}
\includegraphics[scale=0.64]{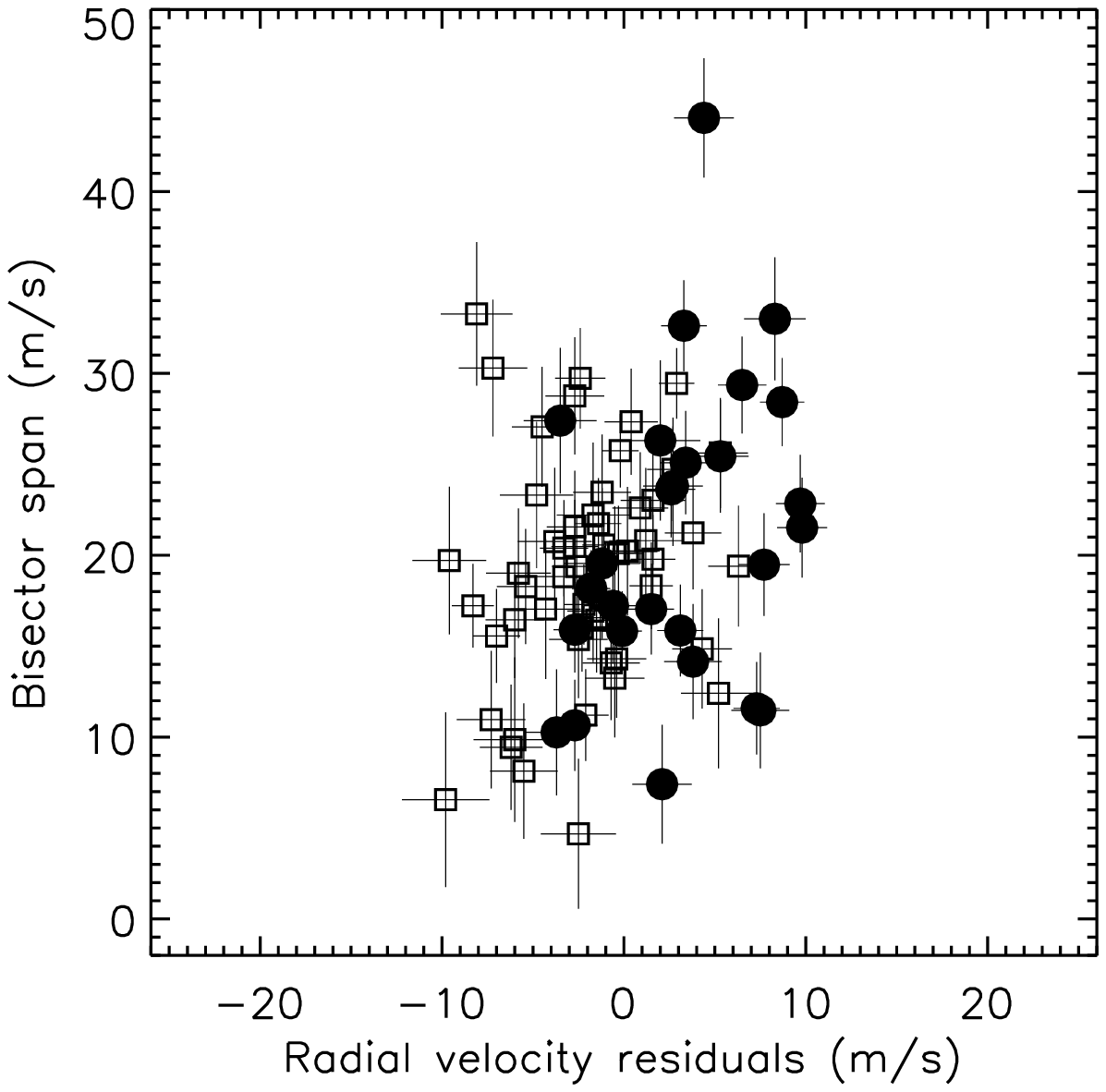}
\caption{Bisector span as a function of the radial velocity (upper panel) and of the residuals of 
the fit with two planets as shown in Figs.~\ref{fig_omc} and~\ref{fig_orb_phas}  (lower panel).
On the lower panel, the filled circles indicate the measurements secured on the 101-day 
interval, where we fitted the stellar jitter (\S~\ref{Fit with stellar activity} and Fig.~\ref{fig_ext}).
The ranges have the same extends in $x$- and $y$-axes on both panels.}
\label{fig_bis}
\end{center}
\end{figure}

The core of the large H \& K \ion{Ca}{ii} absorption lines of \cible\ 
show emissions, indicating chromospheric activity. 
We adopted the averaged value 
$\log{R'_\mathrm{HK}} = -4.8 \pm 0.1$, but it significantly varies with time, between 
extrema $\log{R'_\mathrm{HK}} = -4.68 \pm 0.01$ and $-4.88 \pm 0.01$.
Such stellar activity would imply a significant jitter on the observed stellar radial 
velocities. For a K-type star with this level of activity, Santos et al.~(\cite{santos00}) 
predict a dispersion up to 10~\ms\ for the stellar jitter. This is the order of magnitude 
of the dispersion of our measurements of \cible~(\S~\ref{sect_obs_BD08}).
A classic method to identify stellar activity as the main cause of radial velocity variations
is to look for anti-correlation between radial velocity and the bisector span (see, e.g., 
Queloz et al.~\cite{queloz01}, Boisse et al.~\cite{boisse09}). The upper panel of 
Fig.~\ref{fig_bis} shows  the bisector spans as a function of the radial velocities 
for \cible. As seen above, the bisector is significantly varying; however, no clear 
relations are seen between the two signals. Yet, this technic is less sensitive 
for stars with slow rotations, as this is the case for \cible. Indeed, according to 
Noyes et al.~(\cite{noyes84}) and Mamajek \& Hillenbrand (\cite{mamajek08}), 
the activity level for \cible\ implies a stellar rotation period around~37~days.
The \hip\ data of \cible\ includes  photometric measurements for 63 epochs only; 
this does not allow a rotation period measurement. 

\begin{figure}[h] 
\begin{center}
\includegraphics[scale=0.59]{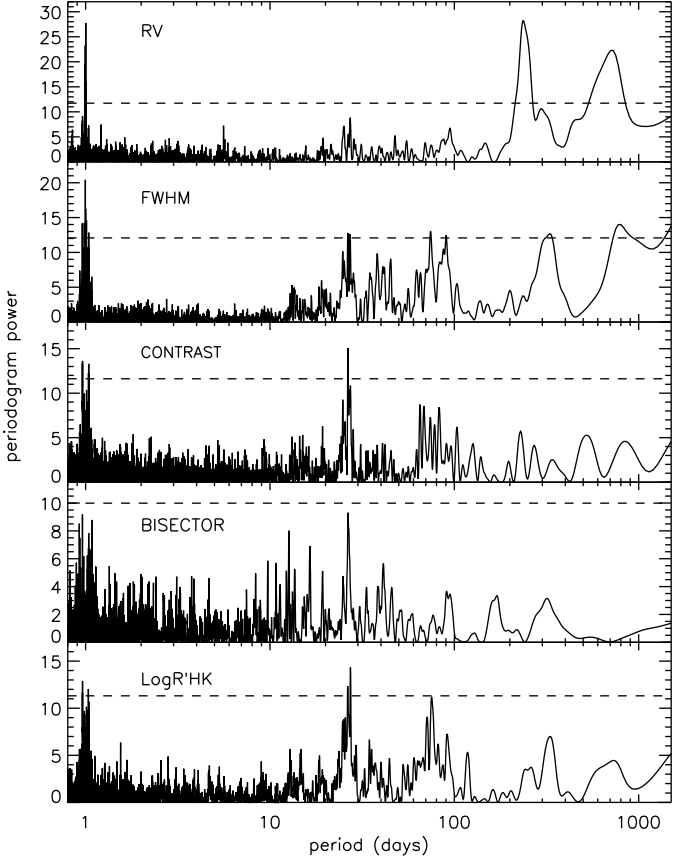}
\caption{Lomb-Scargle periodograms for \cible\ HARPS signals. From top to bottom, 
the five panels show the periodograms of the radial velocities, of the FWHMs, contrasts 
and bisectors of the CCFs, and finally of the $\log{R'_\mathrm{HK}}$ activity indexes. Each of 
these five signals shows a peak in the power around 26~days, which could be interpreted 
as the stellar rotation period. On the upper panel, the radial velocities periodo\-gram shows 
two extra peaks at 5.6 and 238~days, which are not seen in the four other~panels.
The horizontal dashed lines correspond to the false-alarm probability of $1\times10^{-3}$.
}
\label{fig_periodogram1}
\end{center}
\end{figure}

A different estimation of the rotation period could be obtained from the HARPS data through
the stellar activity indicators, as shown in Fig.~\ref{fig_periodogram1}. In this figure are 
plotted the Lomb-Scargle periodograms (Press et al.~\cite{press92}) of five HARPS signals: 
the radial velocities, full widths at half maximum (FWHMs), contrasts and bisectors of the CCFs, 
and of the $\log{R'_\mathrm{HK}}$ 
activity indexes (Table~\ref{table_rv}). 
The fourth last parameters could show signatures of the stellar activity, and in 
particular their modulations with the stellar rotation; the radial velocity variations could as well show 
signatures of the stellar activity and rotation, but also the Doppler reflex motion due to companion(s).
The five periodograms in Fig.~\ref{fig_periodogram1} all clearly show a peak around 26~days.
We interpret it as the signature of modulations of the stellar surface due to activity (flares, spots, 
plages...). This kind of phenomena are expected for such an active star. As there are sporadic 
events altering the surface of the star and having limited life times, they would imply quasi-periodic 
variations of the shapes of the spectral lines, with periods near the stellar rotation period, and 
unconserved phases. The rotation period of \cible\ seems thus to be around 26~days from these 
data. Most of the signals show also a small peak near 13-days, which is the first harmonic of 
the main signal; it is not detected in the radial velocities, however. 
In addition, the five perio\-dograms in Fig.~\ref{fig_periodogram1} show signal between 65 
and 95~days. It may be the signature of the typical duration of cycles of these 
phenomena. 

The radial velocity periodogram show two extra peaks that are not seen for other 
signals in Fig.~\ref{fig_periodogram1}, at 5.6 and 238~days. The fact that these periodic signals 
are detected only in radial velocities suggests it is caused by Doppler reflex motion due to 
companions for \cible\ rather than jitter due to stellar activity. Moreover, those two periods 
do not correspond to any harmonics nor aliases of the 26~days rotation period.
The amplitude of the variations (43~\ms\ peak-to-peak,  see 
upper panel of Fig.~\ref{fig_omc}) 
and the two orbital periods (near 5.6 and 238~days) imply projected masses 
$M_\textrm{p} \sin i$ well below the mass of Jupiter.
The radial velocity variations detected by HARPS would thus originate both from 
stellar activity jitter and planetary companions.

\section{A planetary system around \cible}
\label{sect_planetary_system}

\subsection{Fit without stellar activity}
\label{sect_fit_without_activity}

\begin{figure}[h] 
\begin{center}
\includegraphics[scale=0.44]{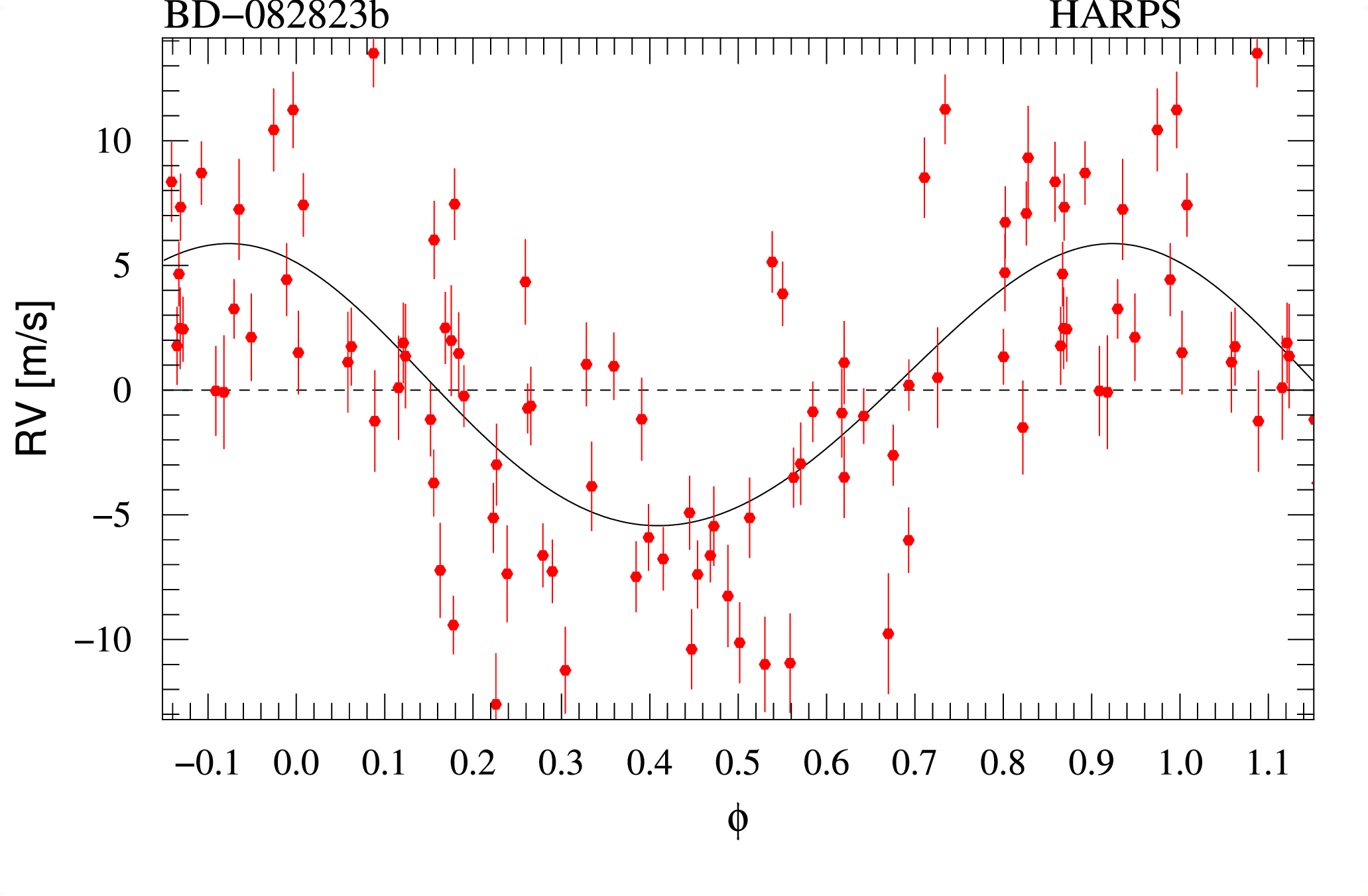}
\includegraphics[scale=0.44]{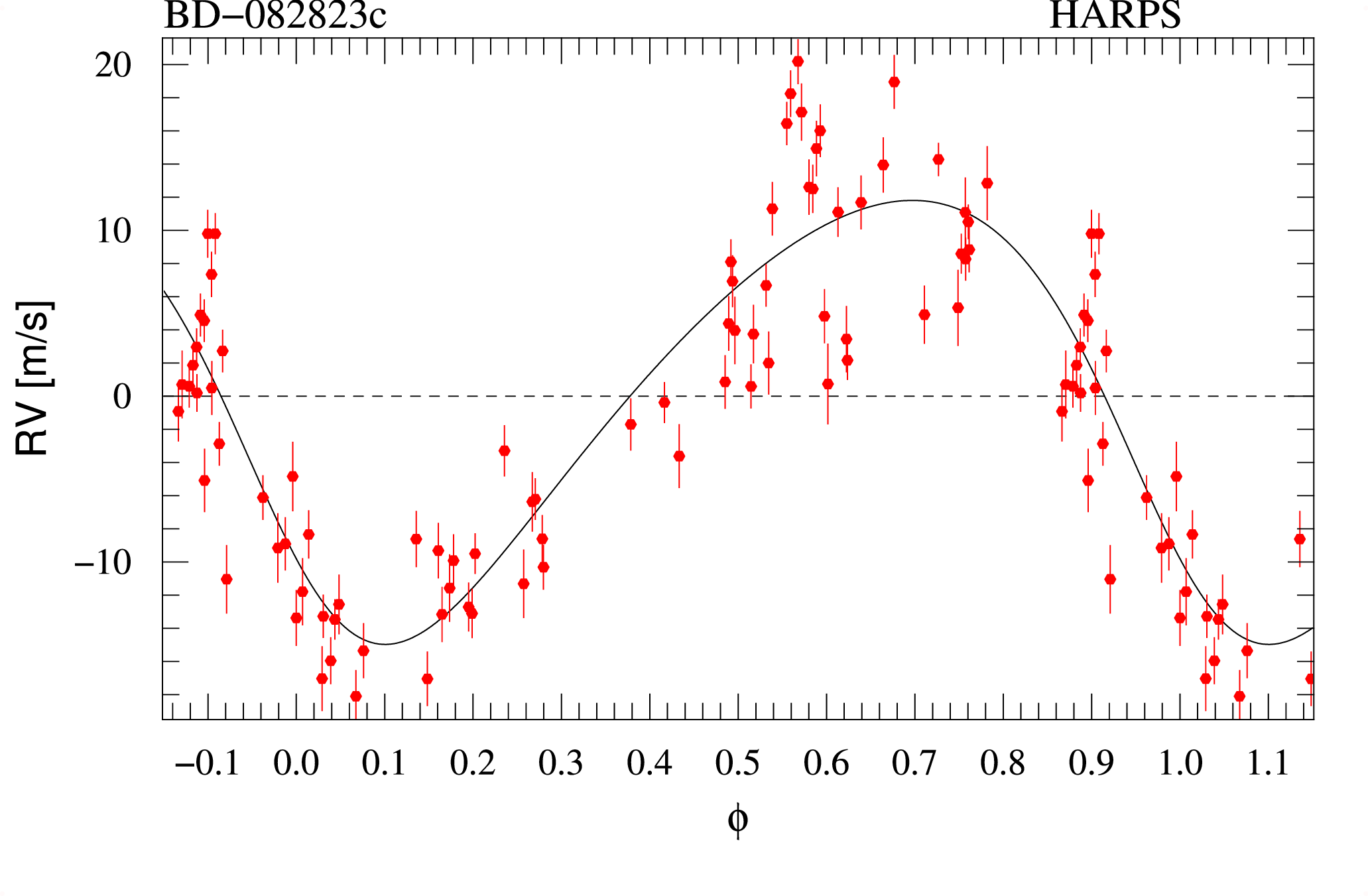}
\caption{Phase-folded radial velocity curves for \cibleb\ ($P=5.60$\,d, top) 
and \ciblec\ ($P=237.6$\,d, bottom) after removing the effect of the other planet.
The HARPS radial velocity measurements are presented with 
1-$\sigma$~error bars, and the Keplerian fits are the solid lines. 
Orbital parameters corresponding to the 
fits are reported in~Table~\ref{table_parameters}.}
\label{fig_orb_phas}
\end{center}
\end{figure}

We first fitted the radial velocities using a two-planet Keplerian model without mutual 
interactions. The results are plotted in Figs.~\ref{fig_omc} and~\ref{fig_orb_phas}. 
The inner planet, \cibleb, produces radial velocity variations with a small 
semi-amplitude $K=6.5$~\ms, corresponding to a planet with a minimum mass 
$M_\textrm{p} \sin i   = 14.4$~M$_{\oplus}$, thus similar to the mass of Uranus. 
Its orbit has a period of $5.60$~days and is circular or with low eccentricity.
The outer planet, \ciblec, yields a larger semi-amplitude, namely $K=13.4$~\ms;
this corresponds to a planet with a minimum mass 
$M_\textrm{p} \sin i = 0.33$~M$_\mathrm{Jup}$, slightly above the Saturn mass. 
The orbital period is $237.6$~days, and the orbit is non-circular ($e=0.19$). 
The derived orbital parameters of the system
are summarized in Table~\ref{table_parameters}.

\begin{table}[h]
  \centering 
  \caption{Fitted orbits and planetary parameters for the \cible\ system, with 1-$\sigma$ error bars.}
  \label{table_parameters}
\begin{tabular}{lcc}
\hline
\hline
Parameters & \cibleb\ & \ciblec\ \\
\hline
$P$ 				[days]			& $5.60\pm 0.02$			&  $237.6\pm 1.5$	\\
$e$								& $0.15\pm 0.15$			&  $0.19\pm 0.09 $	\\
$\omega$ 		[$^{\circ}$]			& $30\pm100$				&  $-233\pm21$	\\
$K$				[\ms]				& $6.5\pm1.0$				&  $13.4\pm1.0$	\\
$T_0$ (periastron)	[BJD]			& $2\,454\,637.7\pm1.6$		&  $2\,454\,193\pm13$ 	\\
$M_\textrm{p} \sin i$	[M$_\mathrm{Jup}$]	& $ 0.045 \pm 0.007$$^\ddagger$  	&  $0.33 \pm 0.03$$^\ddagger$  \\
$M_\textrm{p} \sin i$	[M$_{\oplus}$]		& $ 14.4 \pm 2.1$$^\ddagger$   		&  $104 \pm 10$$^\ddagger$  \\
$a$				[AU]				& $ 0.056\pm 0.002$$^\ddagger$    &  $0.68 \pm 0.02$$^\ddagger$  \\
$V_r$ 			[\kms]			& \multicolumn{2}{c}{$53.345\pm0.001$} 	\\
$\sigma_\mathrm{O-C}$		[\ms]		& \multicolumn{2}{c}{4.3}	 	\\	
reduced \kid						& \multicolumn{2}{c}{3.2}	 		\\ 
$N$								& \multicolumn{2}{c}{83}	 		\\
span 			[years]			& \multicolumn{2}{c}{5.0}                      \\
\hline
\multicolumn{3}{l}{$\ddagger$: using $M_\star =  0.74\pm 0.07$\,M$_\odot$}
\end{tabular}
\end{table}

The reduced \kid\ of the Keplerian fit is 3.2, and the standard deviation of the residuals is 
$\sigma_\mathrm{O-C}=4.3$~m\,s$^{-1}$. This is reduced when compared to the $11$-\ms\ 
dispersion of the original radial velocities, but this is clearly higher than the 1.6-\ms\ typical 
error bars on the individual measurements. The extra dispersion, of the order of 4 \ms, is 
mainly due to stellar activity jitter as seen above. By fitting the two planets without including 
the stellar activity, we assume that the stellar jitter would be averaged out over the five years of 
observations, as activity induces quasi-periodic effects that are not exactly duplicated 
in time 
with the~stellar rotation. This results in a quite large dispersion around the fit with 
respect to the error bars, whereas the periodic signal of the two planets remains coherent 
over the five years of~observations. 

\begin{figure}[h] 
\begin{center}
\includegraphics[scale=0.59]{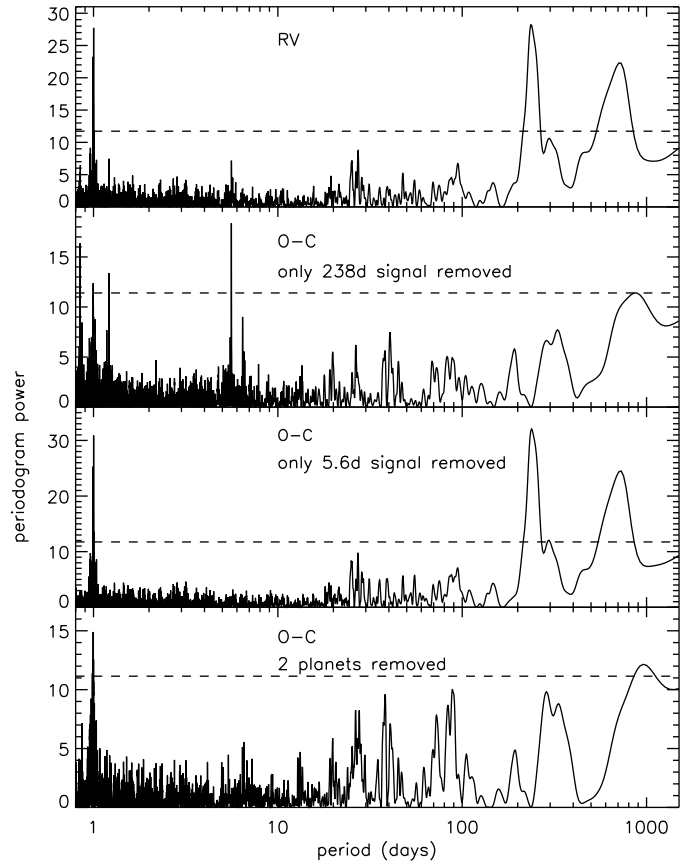}
\caption{Lomb-Scargle periodograms of the HARPS radial velocities. The upper 
panel shows the periodogram computed on the initial radial velocities, without any fit 
removed. The second and third panels show the periodograms computed on the residuals 
of the fits including \ciblec\ or \cibleb\ only, respectively. The bottom panel shows the 
periodogram after the subtraction of the 2-planet fit. 
The horizontal dashed lines correspond to the false-alarm probability of $1\times10^{-3}$.
}
\label{fig_periodogram2}
\end{center}
\end{figure}

Fig.~\ref{fig_periodogram2} strengthens the interpretation in term of planets of the two 
signals at 5.6 and 238~days. The upper panel shows the periodiogram 
of the radial velocities, exactly as the upper panel of Fig.~\ref{fig_periodogram1}. 
On the second panel of Fig.~\ref{fig_periodogram2} is shown the periodiogram of the 
radial velocity residuals, after a fit including \ciblec\ only. The standard deviation of the 
residuals of this fit is 
$\sigma_\mathrm{O-C}=5.9$~m\,s$^{-1}$. On this periodogram, 
the peak at 238~days of course 
is no longer visible. The main peak is this at 5.6~days.
Its false-alarm probability is $5.8\times10^{-5}$.
It is  clearly stronger than the peaks due  to activity.  
We note also the presence of two other peaks, at 0.8 and 1.2~day, that are the 
one-day aliases of the 5.6-day signal in the frequency space ($1\pm1/5.6$). We adopted 
5.6~days as the actual period of this signal instead of one of the two aliases, 
because it is more likely according to the sampling. 
In addition, its peak is higher than those of the two aliases, and Keplerian fits performed with 
0.8 or 1.2~day of orbital periods produce higher residuals dispersion, and eccentric~orbits. 

Similarly, the third panel of Fig.~\ref{fig_periodogram2} shows the periodiogram of the 
radial velocity residuals, after a fit including \cibleb\ only. The peak at 5.6~days is removed 
as well as its aliases, and the 238-day signal remains, clearly above the activity peaks. 
Finally, the bottom panel of Fig.~\ref{fig_periodogram2} shows the periodiogram of the 
residuals after the 2-planet fit shown in Figs.~\ref{fig_omc} and~\ref{fig_orb_phas}. Only 
lower peaks are remaining, most of them being caused by activity. 
Their false-alarm probabilities are larger than of $1\times10^{-3}$.
We note that in addition to the clear peak at one day 
(which is due to the aliases of all the detected signals), all the four panels in 
Fig.~\ref{fig_periodogram2} show a possible peak near 700~days; this could be the signature 
of a third, outer planet. Fits with a curvature in addition to the two planets also suggest a
possible long-period signal (without significant effects on the parameters of 
\cibleb\ and \ciblec). 
Such long-period signals are not strong enough however to claim any detection. Further 
observations of this target on a longer time baseline are mandatory to establish or not 
the presence of the hypothetic planet \cibled. 

The lower panel of Fig.~\ref{fig_bis} shows the bisector spans as a function of the radial velocities 
residuals after the 2-planet fit. This does not show an anti-correlation between those two signals, 
as it could be expected in case of spotted stellar surface due to activity (see, e.g., Melo et 
al.~\cite{melo07}, Boisse et al.~\cite{boisse09}). The absence of anti-correlation, whereas 
activity is supposed to be here the main part of radial velocity residuals, could be due to 
the fact that the rotation velocity is low, or to activity processes more complex than simple
cool stellar spots. It could also be due to the presence of additional companions 
(as the hypothetic \cibled), that~would induce radial velocity variations uncorrelated 
with the~bisectors. 

We note also that we did not find any anti-correlation between the bisector spans and 
the radial velocities with either one or the other planet removed. Such anti-correlation
with just one planet removed would argue that the other signal is due to activity; this is not the case.

\subsection{Fit with stellar activity}
\label{Fit with stellar activity}

In the previous section we have shown a Keplerian fit of the two planets without attempting 
any fit of the stellar activity jitter signal in the radial velocities. Our knowledge of the activity 
is poor; for example the number, locations and sizes of potential stellar spots are far 
to be controlled. We attempt here a naive, phenomenological approach, based on 
the fact that activity shows signals in the periodograms, so stellar jitter has a periodical 
nature that is linked to the rotation of the star. Unfortunately these signals are only 
\textit{quasi-periodical}: they do not reproduce themselves periodically in an identical 
form for a long time, as all these phenomena have limited life times. 
Fitting the stellar jitter with sinusoids 
over the 5-year time span do not provide satisfactory fits. 

\begin{figure}[h] 
\begin{center}
\includegraphics[scale=0.33]{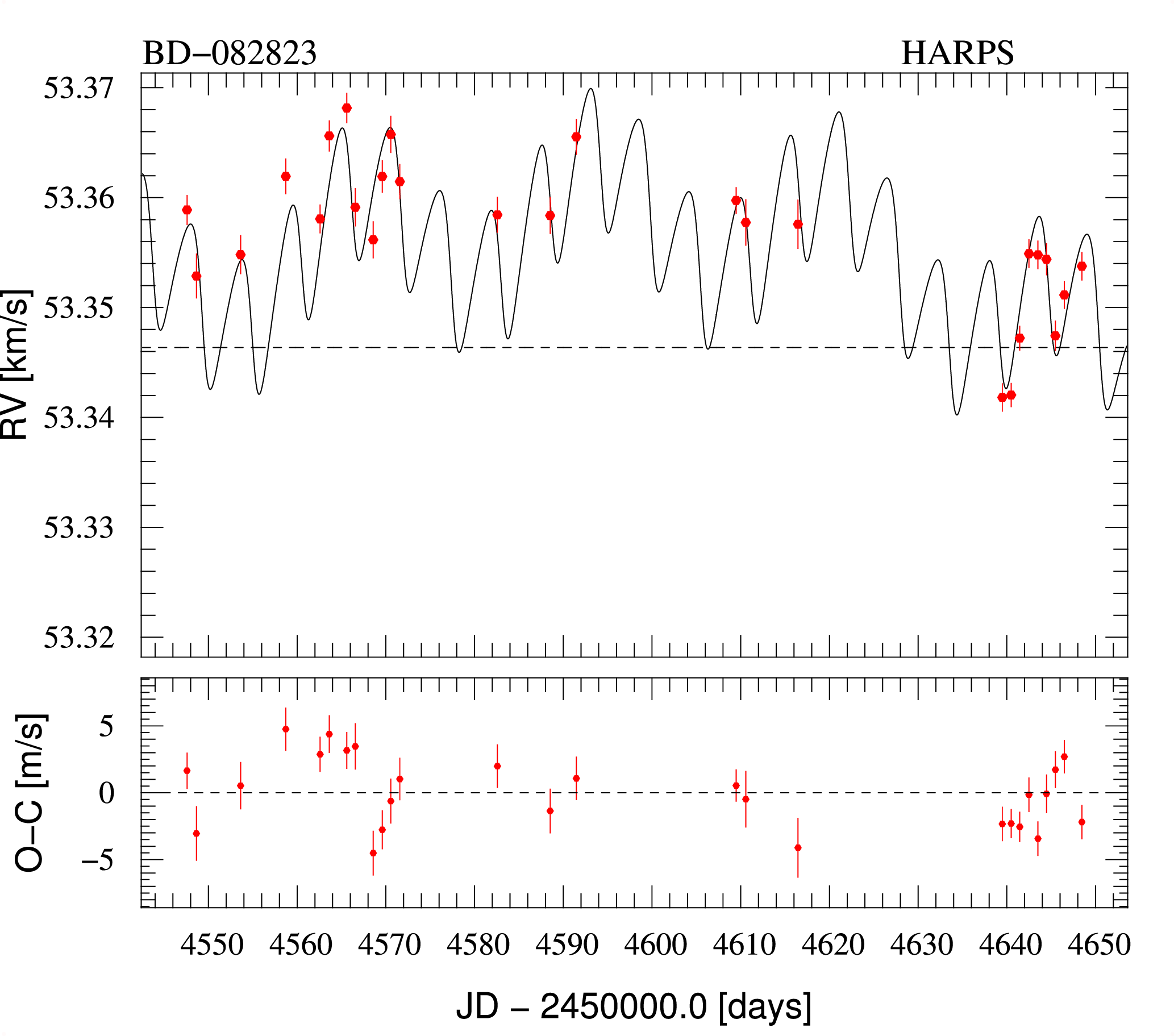}
\includegraphics[scale=0.455]{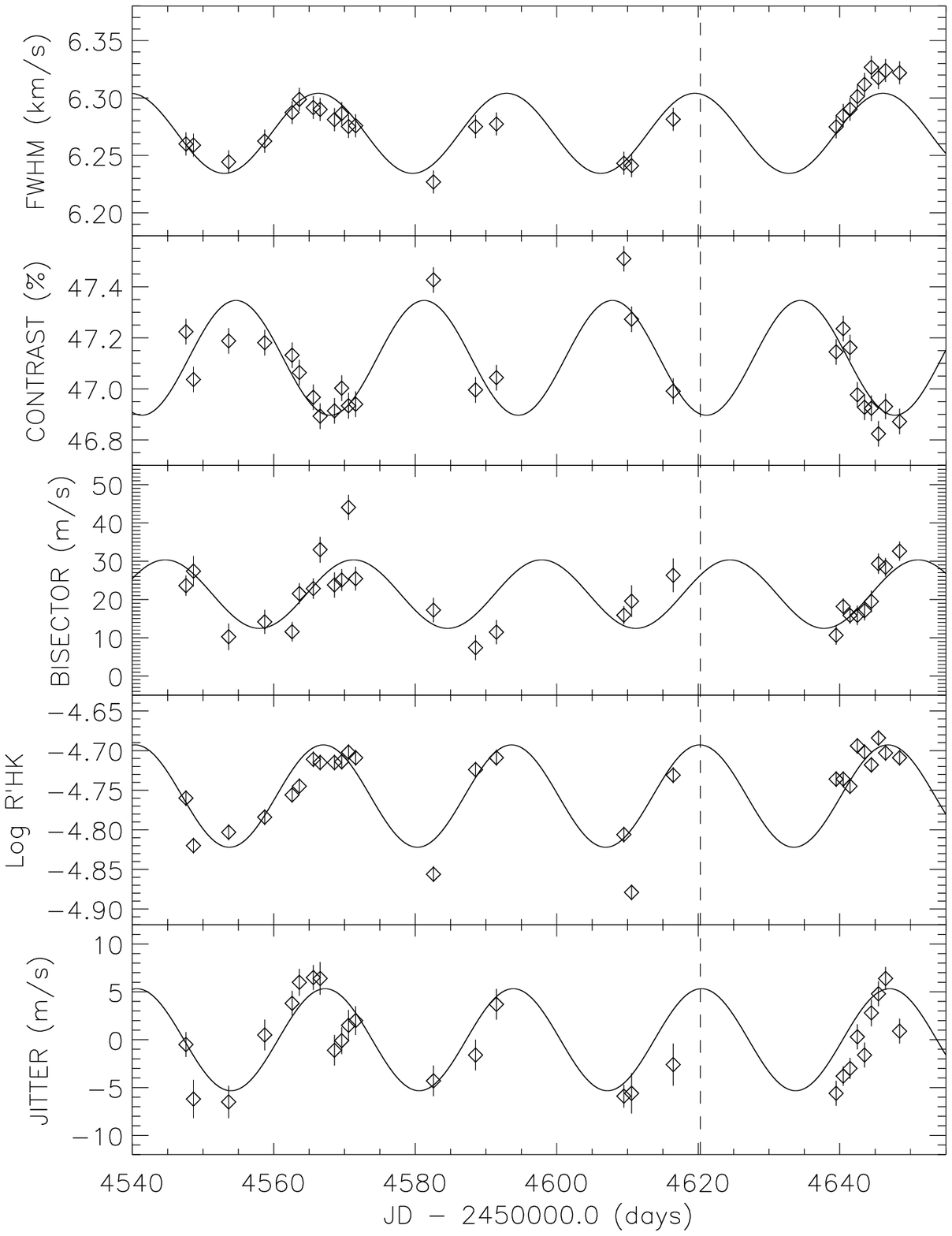}
\caption{Fit of the two planets and the stellar activity on a 101-day interval. 
The \textit{upper plot} shows 
the HARPS radial velocities of \cible\ with a fit including two Keplerian curves for 
the planets and a 26.6-day-period sinusoid for the stellar jitter. 
The \textit{middle plot }show the residuals of this fits. 
The \textit{lower plot} shows on five panels the measurements of the FWHMs, contrasts and bisectors of 
the CCFs, the $\log{R'_\mathrm{HK}}$ and the stellar jitter, respectively from top to bottom. Sinusoid 
fits with 26.6-day period are overplotted. The stellar jitter in the lower panel is the sinusoid that is 
added to the two Keplerian in the upper plot. 
The vertical dashed-line helps visualize the correlations.
Error bars at \1s\ are plotted for all measurements in this figure.
}
\label{fig_ext}
\end{center}
\end{figure}

We thus tried such sinusoid fits on a shorter time span. We chose a 101-day interval, between 
BJD-2\,400\,000 = 54547.5 and 54648.5. This interval has a good sampling (27 
measurements), including two 9-night sequences with almost one measurement 
per night (8 and 9 measurements in the two sequences) at high accuracy (uncertainties 
between 1.0 and 1.7~\ms). This should improve the 
coherence of the stellar jitter in term of periodicity, as well as allowing the 5.60-day-period 
planet to be well sampled. Using sinusoids of periods near 26~days, we fitted 
on this 101-day interval the activity indicators studied in Fig.~\ref{fig_periodogram1}, 
namely the FWHMs, contrasts and bisectors of the CCFs, and the
$\log{R'_\mathrm{HK}}$ indexes. The best solution we obtained was for a 
26.6-day period. Together with the periodograms shown in Fig.~\ref{fig_periodogram1}, 
this allowed us to determine the rotation period: $26.6\pm1.5$~days. 
We fitted on the same interval the radial velocities, 
using Keplerian models for the two planets, and an extra sinusoid to approximate the 
stellar jitter. The parameters of the inner planet were free, as the mass of the outer one. 
The  period, eccentricity and orien\-tation of the orbit of  \ciblec\ 
were fixed to the values obtained above, the 101-day interval being too short to 
constrain this 238-day-period orbit. 

The results of these fits agree with those obtained in \S~\ref{sect_fit_without_activity}.
The fits are plotted in Fig.~\ref{fig_ext}, which shows on the upper panel a 
good match of the radial velocities and the model, especially on the two high-frequency 
observation sequences that show the variations due to \cibleb. 
The lower panels show the activity signals.
The 26.6-day-period sinusoids provide acceptable approximations for their variations, 
which remain however more complex than this simple sinusoid shape. The lower panel shows 
the stellar jitter only, i.e. the sinusoid that is added to the two Keplerian in the upper 
panel. It has an amplitude of $\pm5$~\ms.
The residuals of the radial velocities fit with two Keplerian and a sinusoid is plotted in 
the middle panel of Fig.~\ref{fig_ext}. Its dispersion is reduced 
down to 2.5~\ms, which is an improvement by comparison to the 4.3~\ms\ dispersion 
obtained on the full dataset in \S~\ref{sect_fit_without_activity} without an attempt to 
fit the stellar jitter
(or 4.1~\ms\ dispersion if computed only on the 101-day interval, again without 
stellar jitter fit). 
It remains larger than the error bars, because of the imperfection of 
the sinusoid model we used for the stellar jitter
(and perhaps also because of possible extra planets). 
We restrained this fit to a single 
sinusoid for the stellar jitter, without including extra sinusoids at the periods of 
the stellar rotation harmonics. Indeed, these harmonics are not detected in the 
radial velocities (see \S~\ref{stellar_prop} and Fig.~\ref{fig_periodogram1}).

Despite its simplicity, the simple sinusoids allow correlations to be seen between 
the different parameters on this 101-day time span (Fig.~\ref{fig_ext}, lower panels). 
The FWHMs and contrasts of the 
CCFs are anti-correlated, implying spectral line deformations due to activity that 
let nearly constant their equivalent widths. This could be understood as the 
surface of the CCF is an indicator of the stellar metallicity (Santos et 
al.~\cite{santos02}). 
The velocity jitter is barely correlated with $\log{R'_\mathrm{HK}}$, 
as well as with the FWHM and the bisector of the CCF. The correlation between the 
FWHM and $\log{R'_\mathrm{HK}}$ is different from what was seen for example in 
the case of the spotted star CoRoT-7, for which those two values are rather 
anti-correlated (Queloz et al.~\cite{queloz09}). Also, the apparent correlation 
between $\log{R'_\mathrm{HK}}$ and the radial velocity jitter is different from 
the picture seen in the cases of the active stars 
HD\,166435 (Queloz et al.~\cite{queloz01}), GJ\,674 (Bonfils et al.~\cite{bonfils07}) or
HD\,189733 (Boisse et al.~\cite{boisse09}). 
And finally, there is an apparent correlation between the bisector and the radial 
velocity jitter on this 101-day time span, possibly with a small phase offset 
(see~ Fig.~\ref{fig_ext}). This possible correlation is also shown 
on the lower panel of Fig.~\ref{fig_bis}.
All these relations drawn a picture for \cible\ activity that is quite different from a 
simple scenario where the radial velocity jitter is mainly due to dark spots on the 
stellar surface that modulate the shape of the lines as the star is rotating. Other 
phenomena are likely to occur on this star; they may include pulsations, convections, 
flares, plages, hot~spots...

The effect of the stellar activity on the observed radial velocities of \cible\ is of the 
order of 4~\ms. This is lower than the semi-amplitudes measured for the two detected 
planets, but not negligible. The error bars reported in Table~\ref{table_parameters} 
where derived from \kid\ variations and Monte~Carlo experiments with and without 
stellar activity modeling,  as well as from 
trials and errors with different kinds of sinusoids for the stellar jitter.  
Our poor understanding of the stellar activity of \cible\ put us to remain cautious 
on the obtained error bars. 
The reality 
of the two planets in this range of mass is however well established, their 
signatures being clearly detected at two periods that show no signals linked to 
stellar activity.

\section{Discussion}
\label{sect_conclusion}

Finding transiting planets in the \hip\ epoch
photometry annex does not look promising. 
Our attempt for \textit{a-priori} detections did not succeed, and up to now, only 
two \textit{a-posteriori} detections were performed within the \hip\ data, in the 
cases of HD\,209458b and HD\,189733b that were first revealed from ground
observations. 
Three main limitations make it difficult. First, the error bars on individual 
photometric measurements are of the same order of magnitude 
than the expected signal for transits of giant planets, or even slightly larger. Second, 
the sparse time-coverage allows only a few points to be obtained in a
potential transit. These two limitations make any transit identification at 
the limit of detection. Third, stellar activity could produce false positives, 
which are difficult to identify with the sparse time-coverage. 
The ESA mission Gaia (Perryman et al.~\cite{perryman01})
is awaited as the successor of \hip. Its potential for 
planetary transit discoveries in front of bright stars could be considered, as 
its sensitivity and accuracy would be better than \hip. However, the time coverage 
won't be better than
that of \hip, which will prohibit well-resolved 
light curve studies. Stellar activity should thus also be a limitation for transit 
detections with Gaia. Dedicated surveys have proven to be more efficient 
for transit detections. Ground-based programs allow Jupiter-size planets to 
be detected, with improving accuracies that now point toward planets 
with smaller radii (Bakos et al.~\cite{bakos09}), whereas space-based 
programs allow planets with even smaller radii to be detected (L\'eger et 
al.~\cite{leger09}, Queloz et al.~\cite{queloz09}), as well as planets on longer periods (Moutou et 
al.~\cite{moutou09b}). If accepted, the ESA-proposed space mission PLATO 
(Catala et al.~\cite{catala09})
could permit the detection of such kind of transiting planets in front of 
brighter stars.

If this search in the \textit{Hipparcos} data did not provide any detection 
of new transiting planets, it allowed the serendipitous discovery of a new 
planetary system around \cible. This target was first identified as presenting 
promising radial velocity variations, in which we identified the signature of 
two new extra-solar planets thanks to an intensive monitoring with HARPS.
The inner planet, \cibleb, is a hot-Neptune with a minimum mass of 
14.4~M$_{\oplus}$. Its 5.60-day orbit could be circular or slightly eccentric.
The outer planet, \ciblec, is slightly more massive than Saturn, with a 
minimum mass of 0.33~M$_\mathrm{Jup}$. Its orbit is moderately but significantly 
eccentric, and has a period of 237.6~days. As the masses are low and the 
orbits are distant and nearly circular, the mutual interactions between the 
two planets are negligible. 

The reflex motions that these two planets induce to their host star have 
semi-amplitudes of 6.5 and 13.4~\ms, which can be distinguished from 
the 4-\ms\ jitter due to stellar activity. We summarize here the arguments 
that allow us to conclude that these two signals are due to planets  
and not to stellar activity:
\begin{itemize}
\item
the signals with periods of 5.60 and 237.6~days are seen only in the
radial velocities, and are not seen in the shapes of the lines nor in the 
activity indexes;
\item
these two periods do not correspond to the rotation period of the star, 
nor to its harmonics or its aliases;
\item
with the available data, the signal is coherent over the 5-year time span
of the observations.
\end{itemize}
The radial velocities can be fitted 
on a short time span
by two Keplerian and a sinusoid 
that models the stellar activity.
Acknowledging that we do not have a good understanding of the stellar 
activity processes, we can alternatively fit the radial velocities
using two Keplerian only. By doing that we assume that the stellar jitter 
is damped and averaged out over the 5-year time-span, and would only 
lead to a dispersion of the residuals of the fit that are larger than expected 
from the accuracy of the radial velocity measurements. 
Additional planetary companions are possible in this system, so 
the monitoring of \cible\ should be continued.

Most of the Neptune-mass and Super-Earth planets found by the HARPS 
GTO program are detected from a survey of about 200 non-active solar-type 
stars. They are monitored with numerous high-precision measurements, using
simultaneous thorium calibration and exposures of 15-minute duration to 
average out the stellar oscillations (Santos et al.~\cite{santos04b}, 
Udry et al.~\cite{udry06}, Lovis et al.~\cite{lovis06}, Bouchy et al.~\cite{bouchy09}, 
Mayor et al.~\cite{mayor09}). This is not the case for \cibleb, that was 
discovered from lower signal-to-noise exposures, without simultaneous 
calibration, and as a serendipitous result of a program that preferentially 
selected active stars. \cible\ is one of the most active stars around which a 
low-mass planet have been found, together with CoRoT-7 
(Queloz et al.~\cite{queloz09}) and a few M-dwarfs (Bonfils et al.~\cite{bonfils07}, 
Forveille et al.~\cite{forveille09}). 
Those cases support the fact that active~stars should not be neglected in exoplanet 
hunts, even when searching for low-mass planets.

\begin{figure}[h] 
\begin{center}
\includegraphics[scale=0.52]{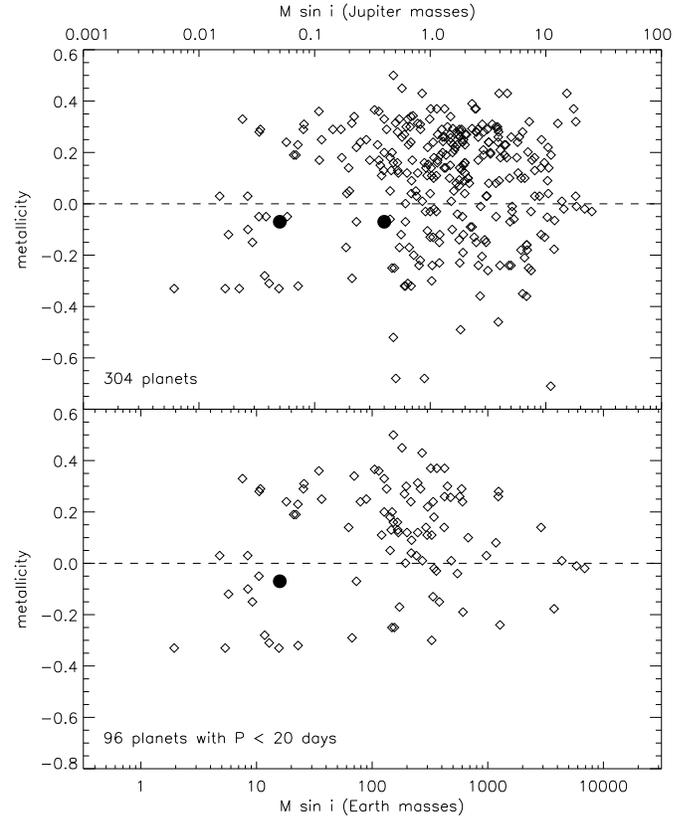}
\caption{Stellar metallicities as a function of the planetary masses, for 304 known extra-solar 
planets (\textit{top}), and for the 96 ones among them with orbital periods shorter than 
20~days (\textit{bottom}). \cibleb\ and \ciblec\ are indicated by the filled circles. Giant 
planets are preferentially hosted by stars with higher metallicities than low-mass planets.}
\label{fig_metal}
\end{center}
\end{figure}

\cible\ has a metallicity that is slightly lower than that of the Sun. This strengthens 
the fact that whereas giant planets are more frequently found around stars with 
super\-solar metallicities, this tendency is not seen for 
Neptune-mass nor Super-Earth planets.
This is shown in Fig.~\ref{fig_metal}, that presents the minimum masses of known 
exoplanets as a function of the metallicity of their hosting stars. The data are taken from 
the compilation of the Extrasolar Planets Encyclopedia\footnote{http://exoplanet.eu}.
While planets with masses larger than 1~M$_\mathrm{Jup}$ are more numerous around 
over\-metallic stars, there are clearly more numerous around under\-metallic stars for 
planetary masses below 20~M$_{\oplus}$. 
The same effect is seen if we select only the 
planets with orbital periods of 20~days or shorter, i.e. in the period regime where most 
of the low-mass planets are detected up to now (lower panel of Fig.~\ref{fig_metal}).
Indeed, among the $\sim30$ planets with a projected mass lower than 
0.1~M$_\mathrm{Jup}$, only four have orbital periods longer than 20~days (namely 
Gl\,581d, HD\,40307d, HD\,69830c and d).

Planet formation models based on the core accretion scenario (Mordasini et al.~\cite{mordasini09a}, 
\cite{mordasini09b}) have shown that the sharply rising probability of detecting 
giant planets with stellar metallicity can be at least partially accounted for by the fact 
that metal rich systems favor the formation of massive planets and that the radial 
velocity technique is most sensitive to massive bodies. Observations also seem to 
indicate that this correlation vanishes for Neptune-mass planets, a trend also found 
in population synthesis calculations (Mordasini et al, in preparation). These calculations 
even show that the correlation reverses for very small planetary masses 
($< 3-5$~M$_{\oplus}$), a prediction that will have to be confirmed by observations. 
In fact, a careful analysis of the theoretical models indicate that the critical parameter 
is the overall mass of heavy elements rather than the metallicity. Since this mass is 
determined by the metallicity and the mass of the initial proto-planetary disk (which is 
unknown), it makes a straight interpretation more difficult. For example, models predict 
that low mass objects orbiting metal rich stars or relatively massive planets orbiting 
metal poor stars are also possible albeit they should be rare. 

The \cible\ system also agrees with the tendency of low-mass planets to be preferentially 
found in multiple systems. Almost 70~\%\ of the $\sim30$ planets with projected masses
lower than 0.1~M$_\mathrm{Jup}$ are found in multiplanet systems, whereas this ratio 
is only $\sim25$~\%\ for all the $\sim350$ known planets. Considering again only the known
planets with orbital periods shorter than 20~days, less than $\sim20$~\%\ of them are found 
to be in multiplanet systems: those $\sim20$~\%\  are mainly low-mass~planets.

No photometric search for transits have been managed for \cible\ from 
follow-up ground-based observations; depending on the unknown inclination 
$i$ of the orbit, the transit probability for \cibleb\ is about 9~\%. Its expected depth 
is out of reach of the \hip\ photometric accuracy.

\begin{acknowledgements}
We would like to thank F.~Pont, F.~Arenou, I.~Boisse, X.~Bonfils, 
J.-M.~D\'esert, R.~Ferlet, J.~Laskar, D.~Sosnowska and A.~Vidal-Madjar 
for helps and discussions, as well as the different observers from other 
HARPS programs  who have also measured \cible.
We are grateful to the ESO staff for their support on the HARPS instrument.
GH, FB and ALDE acknowledge support from
the French National Research Agency (ANR-08-JCJC-0102-01).
DE acknowledges financial support from the CNES.
NCS would like to thank the support by the European Research Council/European Community under the FP7 through a Starting Grant, as well from Funda\c{c}\~ao para a Ci\^encia e a Tecnologia (FCT), Portugal, through programme Ci\^encia\,2007, and in the form of grants reference PTDC/CTE-AST/098528/2008 and PTDC/CTE-AST/098604/2008.
\end{acknowledgements}

\appendix

\section{Selection of targets for transiting planets search in the \hip\ database}
\label{appendice_1}

We present in this Appendix the systematic search we 
managed in the \hip\ epoch photometry for planetary transit 
candidates (\S~\ref{sect_hipparcos}).

First, we selected 23\,304 stars among the 118\,204 of the \hip\ epoch
photometry catalog, according to the following~criteria:
\begin{itemize}
\item
$B-V>0.4$ (F2 type and later);
\item
$\pi > \sigma_\pi$ (defined parallax);
\item
empty H48 field (reference flag for photometry);
\item
H52 field different from D, P, or R (variability types);
\item
$R_\star < 2 \, {\rm R}_\odot$ (small stellar radius).
\end{itemize}

Then we kept only targets with at most one epoch brighter than
$3\,\sigma$ from the average magnitude (in order to remove remaining
variables) and at least 40 different available epochs. We performed
the periods' research on the 17\,800 remaining stars.

For each target, we took the two faintest epochs and we scan all the
possible periods in order to have those two points in a transit. Then
we adopted the period producing the lowest flux in the drop, and we
quantified this solution with the parameter $\alpha$, defined as:

$$\alpha = \frac{<V_{\rm in}> - <V_{\rm out}>}
{\sqrt{\sigma_{<V_{\rm in}>}^2 + \sigma_{<V_{\rm out}>}^2}}$$

\noindent
where $<V_{\rm in}>$ and $<V_{\rm out}>$ are the averaged magnitudes
of the epochs respectively in and out the drop, and $\sigma$ the
standard deviations. The higher $\alpha$ is, the deeper and more
significant the drop is.
We also computed the $\chi^2$ for the fit of the epochs in the drop
with a transit curve for a planet with a radius $R_{\it p} = (0.11 \pm
0.04)\, {\rm R}_\odot$, implying a $\Delta V_{0.11}$ drop in magnitude:

$$\chi^2 = \sum_{\rm transit} \frac{(V_i - <V_{\rm out}> -
\Delta V_{0.11})^2} {\sigma_{V_i}^2 + \sigma_{<V_{\rm out}>}^2 + 
\sigma_{\Delta V_{0.11}}^2}.$$

We finally computed the goodness of fit (gof) for this $\chi^2$.  The
selection of candidates is based on those two parameters: $\alpha$ 
and the goodness of fit.

\begin{figure}[h] 
\begin{center}
\includegraphics[height=7.7cm,angle=-90]{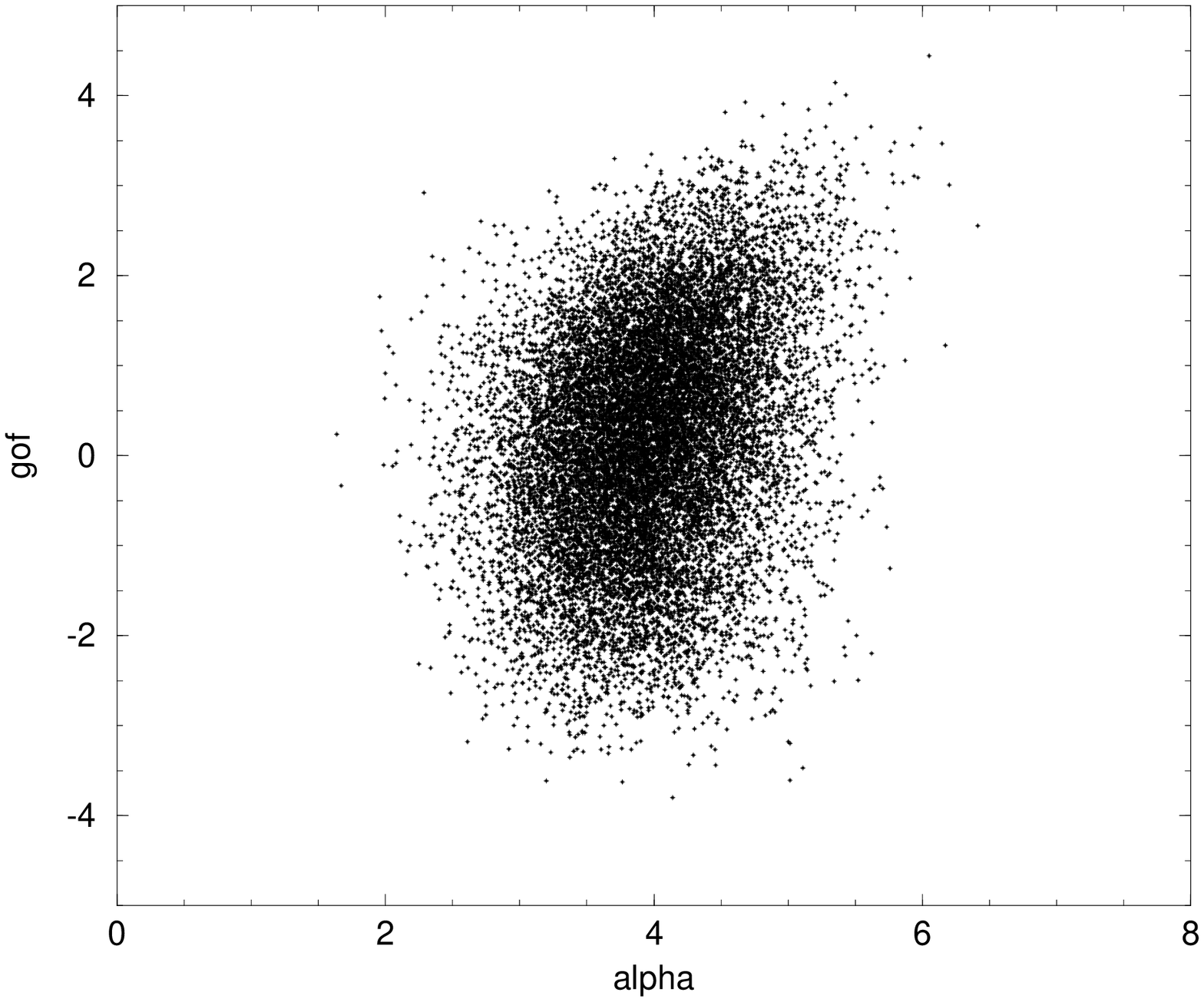}
\includegraphics[height=7.7cm,angle=-90]{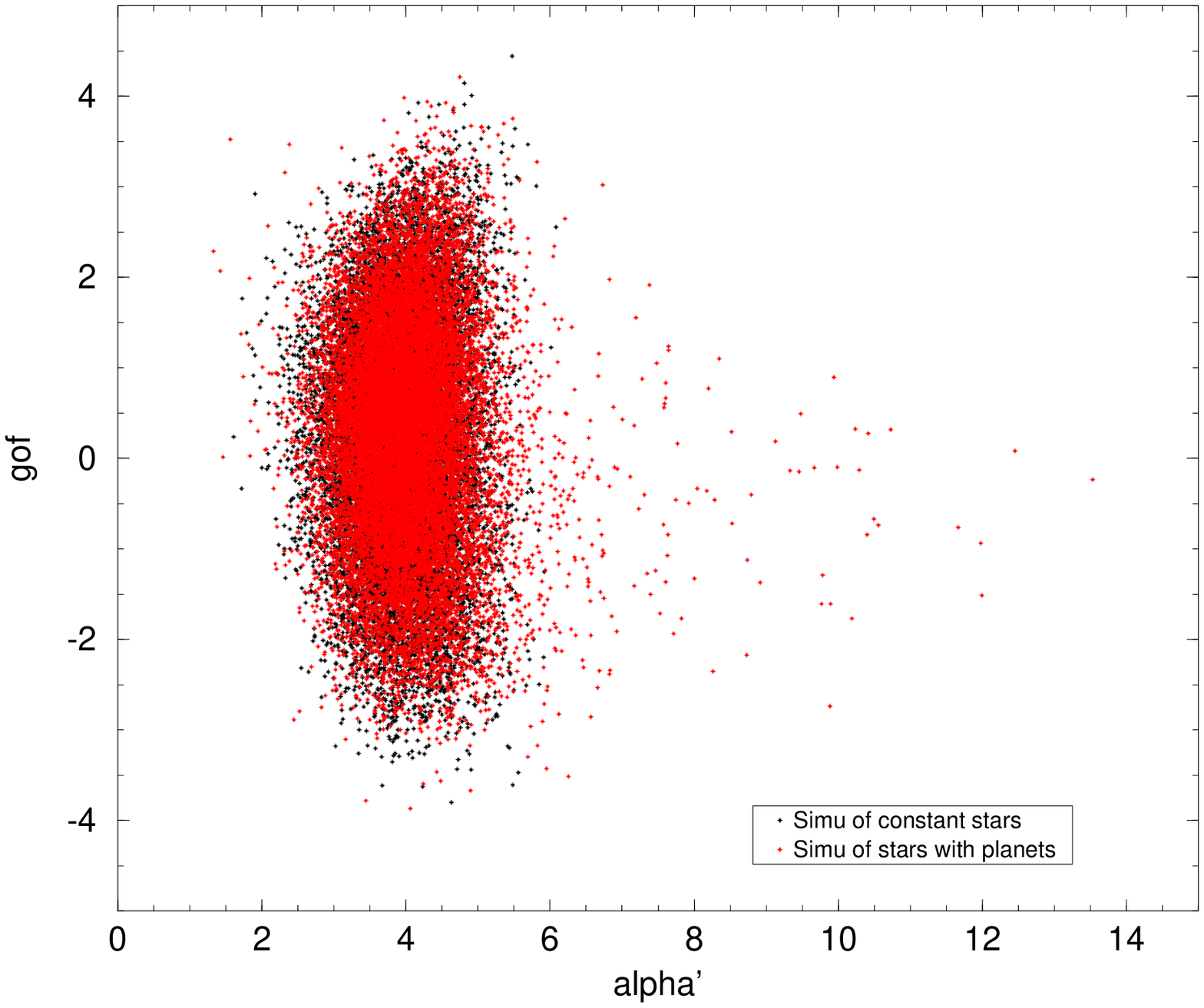}
\includegraphics[height=7.7cm,angle=-90]{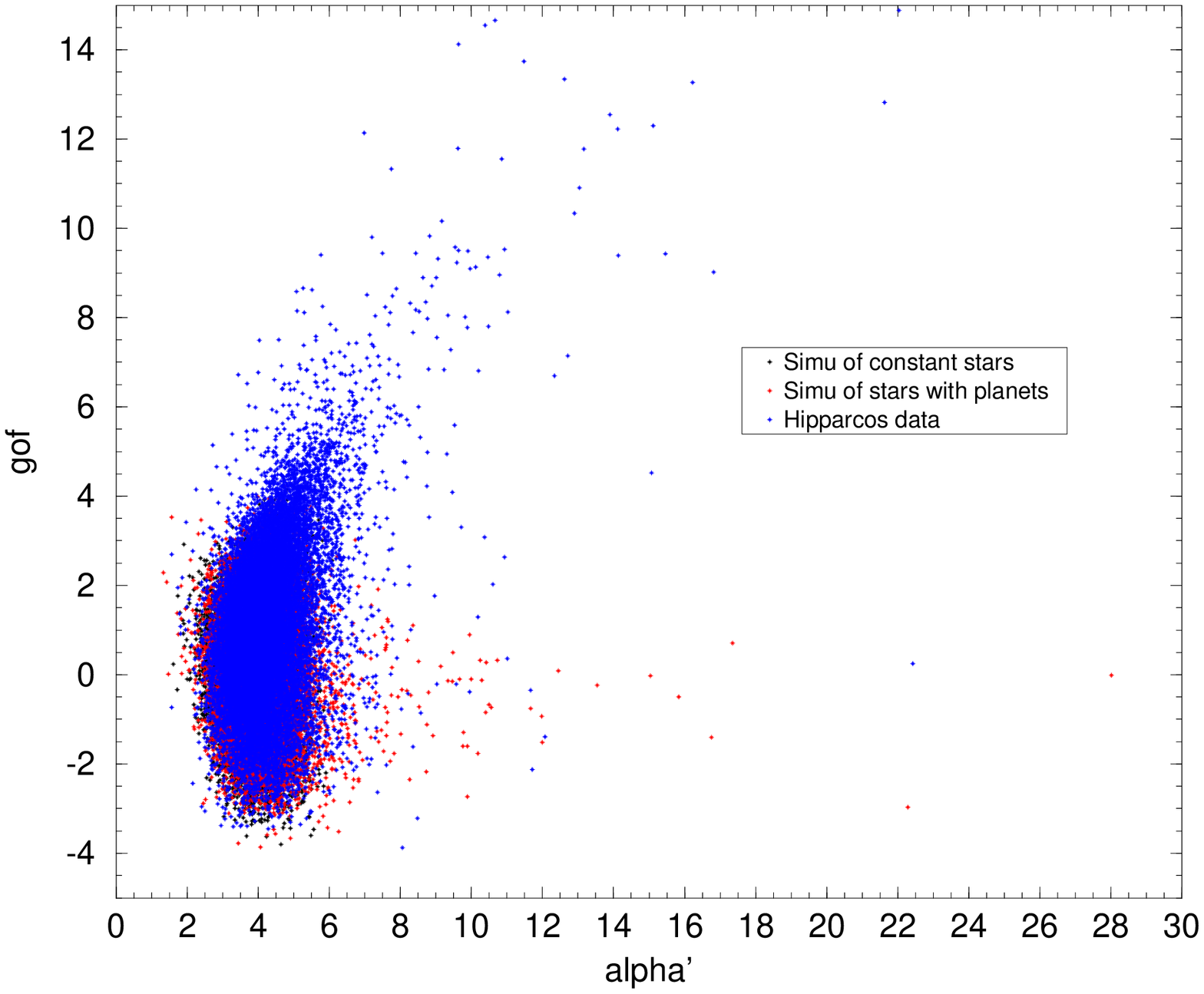}
\caption{Threshold tuning simulations. 
The parameters $\alpha$ and $\alpha'$ characterize the depth 
and the significance of the transit, and \textit{gof} the goodness of fit.
These parameters are defined in the text.
\textit{Top}: Constant stars.
\textit{Middle}: Constant stars (black) and stars with transits (red).
\textit{Bottom}: Actual data (blue) and simulations for constant stars (black)
and stars with transits~(red).
}
\label{fig_simu}
\end{center}
\end{figure}

We performed two simulations in order to tune the threshold: for the
17\,800 stars selected above, we assumed in the first simulation that
all the stars are constant, whereas in the second simulation, we
assumed that all the stars host a transiting planet with a radius
chosen randomly within $0.07-0.15 \, {\rm R}_\odot$. We kept the
individual error on the photometry on each epoch.

The upper panel of Fig.~\ref{fig_simu} shows the goodness of fit as 
a function of $\alpha$ for the first simulation. The distribution is skew 
as $\alpha$ and the goodness of fit are correlated.  In order to make easier the selection, 
we fitted the distribution with a line, and computed $\alpha'$, which is 
corrected (with a first-order polynomial) from this skew. The middle panel 
of Fig.~\ref{fig_simu} shows the goodness of fit as a function of $\alpha'$
for the two simulations. $\alpha' > 5.3$ appears to be an appropriate
threshold for transit candidates selection: indeed, 127 and 489 stars
are above this limit in the first and second simulation,
respectively. We can thus estimate to $(489-127)/17800=2$\% the
detection rate.  With 0.06\% chance that a given star harbors a
transiting extra-solar planet, the 17800 targets of our latest sample
should include $\sim11$ transiting planets. The expected transiting
planet detection number with our method should thus be slightly less
than unity ($11\times2\%=0.2$).

The actual data for the 17800 selected targets are plotted in the lower 
panel of Fig.~\ref{fig_simu}. The stars with high $\alpha'$
and high goodness of fit are eclipsing binaries that were not identified 
in the \hip\ catalog. In order to remove those binaries, we removed targets 
with goodness of fit larger than 2.5. We also removed targets with 
$R_*>1.3\,{\rm R}_\odot$, and obtained a list of
candidates, sorted by decreasing $\alpha'$.  

Targets referenced in SIMBAD as binaries, active, or variable stars
were removed from the obtained list, as well as targets with already known
planets at the time of the observing run (HD\,70642, HD\,39091, HD\,10647, 
HD\,4208, HD\,17051, HD\,13445, HD\,75289, and HD\,83443 -- we checked 
that the \hip\ photometric data did not include periodic variations at their
periods). 

Using HARPS, we performed follow-up observations of 194 of these selected 
targets in December~2004, as part of the program 074.C-0364. 
The apparent magnitudes of the observed stars range from
4.9 to 11.5.  Exposures of typically a few minutes duration were obtained, 
allowing 70 to 90 targets to be observed each night. Errors on the measured 
radial velocities are typically of the order of~2\,m/s.
In order to identify hot Jupiters,~targets were removed from the
candidate list after one, two, or three observations, according to the
following observational~strategy:

\begin{itemize}
\item
observations were stopped after one HARPS measurement if:
1) there are two peacks in the CCF; only 
one SB2 was found during our observations (HD\,23919), binaries having 
been previously removed from the candidate list from \hip\ flags and 
SIMBAD checks;
2) the full width at half maximum of the CCF is larger than 15\,km/s, 
which prohibit accurate radial velocity measurements (17 stars);
\item
observations were stopped after two HARPS measurements if:
1) the radial velocity difference $\Delta \mathrm{RV}$ between the two 
measurements is larger than 2.5\,km/s (no SB1 were found);
2) $\Delta \mathrm{RV}<5$\,m/s (113 stars);
\item
observations were stopped after three HARPS measurements if
$\Delta \mathrm{RV}<20$\,m/s (26 stars).
\end{itemize}

Radial velocity variations larger than 20~\ms\ were measured for 37 stars,
i.e. 19\% of our observed sample. However, those variations are mainly caused by stellar 
activity (see \S~\ref{sect_hipparcos}).

\end{document}